\documentstyle[twocolumn,aps]{revtex}
\begin{document}
\title{Confined magnetic guiding orbit states}
\author{J. Reijniers, A. Matulis,\cite{matulis} K. Chang,\cite{chang} F.
M. Peeters\cite{peeters}}

\address{Departement Natuurkunde, Universiteit Antwerpen (UIA), \\
Universiteitsplein 1, B-2610 Antwerpen, Belgium}
\author{P. Vasilopoulos}
\address{Department of Physics, Concordia University, \\
1455 de Maisonneuve Ouest, Montr\'{e}al, Qu\'{e}bec Canada, H3G1M8}

\address{}
\address{\mbox{}}
\address{\parbox{14cm}{\rm \mbox{}\mbox{}\mbox{}
We show how snake-orbit states which run along a magnetic edge can
be confined electrically. We consider a two-dimensional electron
gas (2DEG) confined into a quantum wire, subjected to a strong
perpendicular and steplike magnetic field $B$/$-B$. Close to this
magnetic step new, spatially confined bound states arise as a
result of the lateral confinement and the magnetic field step.
The number of states, with energy below the first Landau level,
increases as $B$ becomes stronger or as the wire width becomes
larger. These bound states can be understood as an interference
between two counter-propagating one-dimensional snake-orbit
states. \ \\}}
\address{\mbox{}}
\address{\parbox{14cm}{ \rm PACS 73.20.-r, 73.23.-b, 73.40.-c}}

\maketitle
Recently there has been a growing interest in
the study of the influence of inhomogeneous magnetic fields on the
electronic properties of quantum structures. The application of
magnetic fields to otherwise pure electrical systems opened the
door to a whole new area of physics. The confining role of
the magnetic fields can indeed alter the transport properties of, e.g., a
two-dimensional electron gas (2DEG) to a measurable extent and therefore possibly result in
useful magnetoelectronic devices.\cite{peeters99}

With regard to localized electron states, the confinement was
previously realized essentially by application of pure electric
potentials or pure inhomogeneous magnetic fields. However, the
main problem with the latter method is of practical nature: it is
very hard to create an inhomogeneous magnetic field which is
strong enough to confine electrons such that the resulting energy
quantization is large enough to detect.\cite{reijniers01b,kim01}
In contrast to previous approaches, we propose a system which
combines these two different ways of confinement, magnetic and
electric, and which can be realized experimentally.

We consider the so-called snake-orbit states, which have been
studied intensively in the recent
past.\cite{reijniers00,nogaret00,reijniers01} These states are
localized magnetically in one direction by use of large magnetic
field gradients; a steplike magnetic field profile, which varies
abruptly $B$/$-B$, e.g., along the $x$ direction, is a typical
example, experimentally realizable. A typical feature of these
states is that they propagate perpendicular to the magnetic field
discontinuity (along the $y$ direction) and exhibit a large
mobility.\cite{nogaret00} As we show below, reducing this degree
of freedom with hard-wall confinement leads to their complete
spatial confinement.

To start with, we consider electrons in the $(x,y)$ plane
subjected to a perpendicular magnetic field ${\mathbf
B}=[0,0,B(x)]$ which changes sign at $x=0$ such that
$B(x,y)=B$sgn($x$). Along the $y$ direction the electrons are
confined into a quantum wire by a hard-wall potential
$V(|y|>{W})=\infty$, else $V(y)=0$, where $W$ is the width of the
wire. We solve the two-dimensional (2D) stationary
Schr\"{o}dinger equation

\begin{equation}
\{H-E\}\Psi (x,y)=0 \label{hamil}
\end{equation}
with the dimensionless Hamiltonian

\begin{equation}
H=-[\partial ^{2}/\partial x^{2}+ (\partial /\partial y+i|x|B)
^{2}]/2+V(y),
\end{equation}
where $|x|B$ is the $y$-component of the vector potential
$\mathbf{A}$. The coordinates are measured in units of $W$,
energy is in units of $E_{0}=\hbar ^{2}/m_{e}W^{2}$, the magnetic
field in $B_0=\hbar /eW^2$. $l_{B}= \sqrt{\hbar
/eB}=W\sqrt{B_0/B}$ is the magnetic field length. For a typical
GaAs-heterostructure with $W=5000$ \AA, we obtain $E_0 \approx
0.0056$ meV and $B_0 \approx 26.4$ Gauss. The magnetic length
reads $l_B\approx 0.045$ $W$ = 224 \AA.

This quantum mechanical problem cannot be solved analytically.
One way to solve it is by using the mode-matching technique as in
Ref.~\onlinecite{laux88}. In the present work we follow a
different approach and expand the wavefunction in basis functions
of a 2D box in the absence of a magnetic field with boundaries at
$|x|=L/2,$ $|y|=W/2,$ and the length of the wire $L$ chosen
sufficiently large such that it has no influence on the energy of
the bound state. Thus,
\begin{equation}
\Psi (x,y)=\sum_{n=1}^{\infty}\sum_{m=1}^{\infty }c_{nm}\phi
_{nm}(x,y), \label{sumnum}
\end{equation}
is the general form of the wave function and
\begin{equation}
\phi_{nm}(x,y)=\frac{2}{\sqrt{WL}}\sin
[\frac{n\pi}{L}(x+\frac{L}{2})]
[\frac{m\pi
}{W}(y+\frac{W}{2})]
\end{equation}
is the basis function of order $n,m$. Inserting
Eq.~(\ref{sumnum}) into Eq.~(\ref{hamil}) we obtain the secular equation

\begin{equation}
|H_{ij}-E\delta _{ij}|=0, \label{secular}
\end{equation}
where $H_{ij}=\langle\Psi _{i}|H|\Psi _{j}\rangle$ is the complex
matrix element of the Hamiltonian operator with $i=(n,m)$. The
energy of the bound state is then determined by solving Eq.
(\ref{secular}). In principle an infinite number of basis
functions is required to describe the wavefunction and hence to
calculate the exact energy; in practice, though, we found that
$n,\,m\approx 50$ terms are sufficient in order to determine the
energy with an accuracy of $10^{-4}$.

If $B$ is larger than zero, an electron can be bound to the
magnetic edge. We look at relatively large magnetic fields since
we are interested in snake-orbit states, which are rather fragile
structures and show themselves only for weak confining
potentials. That is why we consider the case when the width 
of the wire $W$ exceeds the magnetic length $l_B$. The discrete
energies in the box with dimensions $(L,W)$ subjected to a
homogeneous magnetic field (dots) together with the steplike
magnetic field profile (open dots) with strength $B/B_0=500$ are
shown in Fig.~\ref{fig:output}. For the homogeneous magnetic
field we clearly distinguish the first Landau level (solid dots).
This level is also clearly seen for the $B$/$-B$ profile (open
dots) and corresponds to electrons in cyclotron orbits on the
left or right of the magnetic edge. But in addition, we now
observe states with energy lower than the first Landau level.
Inspection of the wavefunctions, plotted in the inset of
Fig.~\ref{fig:output} for $N=1$ and $N=5$, shows that the
corresponding states are bound close to the magnetic boundary
(situated at $x=0$).

The energy of the bound states, below the first Landau level, is
shown in Fig.~\ref{fig:energy1} as a function of the magnetic
field strength $B/B_{0}=W^2/l_B^2$ in units of $\hbar \omega_c$,
where $\omega_c=eB/m_e$ is the cyclotron frequency. We notice that
with increasing $B$ the ground state decreases in energy and
approaches the limiting value $E/\hbar\omega_c=0.295$ for
$B\rightarrow \infty$. As $B$ increases the number of these bound
states increases.

Actually these new bound states can be interpreted as an
interference between two states propagating along the magnetic
edge, i.e., perpendicular to the wire. These states have the same
energy, lower than the first Landau level, but are propagating in
opposite directions with different momenta, and for some energies
they form a standing wave.

This can be better understood if we first look at the situation
without confinement along $y$ described by $V(y)=0$, that was
discussed in Ref.~\onlinecite{reijniers00}. In this case one has
to solve Eq.~(\ref{hamil}) with $W\to\infty$; now all lengths are
expressed in units of $l_B$ and the energy in units of $\hbar
\omega_c$. Writing the wave function as

\begin{equation}
\psi(x,y)=e^{-iky}\phi_{n,k}(x)/\sqrt{2\pi},
\end{equation}
where $k$ is the wave vector in the $y$ direction, the problem
reduces to solving the 1D Schr\"{o}dinger equation

\begin{equation}
(1/2)[- d^2/dx^2+(|x|B-k)^2]
\phi_{n,k}(x)=E_{n,k}\phi_{n,k}(x),
\end{equation} which we do numerically.

The energy spectrum, plotted in Fig.~\ref{fig:spectrum} as a
function of $k$, shows that to every energy $E\leq \hbar
\omega_c/2$ there correspond two states which propagate along the
magnetic boundary with opposite velocities $v_y$ ($v_y=-\partial
E/\partial k$). The $v_y>0$ states are the classical snake-orbit
states while those with $v_y<0$ do not have any classical analog.

In order to understand the nature of these different states, we
distinguish two different regions, A and B. In region B the
electron cyclotron orbit intersects the magnetic boundary and the
electron motion can be understood classically. In region A there
is no classical analog for the quantum mechanical electron
propagation: classically, the orbit does not reach the magnetic
boundary and the electron cycles around in closed orbits.

Since the lowest energy is at the border between the classical
and non-classical region, only one of the two states with the
same energy $E<\hbar\omega_c$ can be understood classically, the
other cannot. Nevertheless, one could comprehend propagation of
this state more or less, by looking at
Fig.~\ref{fig:trajectories}, where a schematic picture is drawn,
together with the corresponding wavefunctions and their effective
potential. If one adds tunneling to the classical picture, such
that an electron can tunnel from a cycling state on the left to a
cycling state on the right, and vice versa, one can imagine an
electron to propagate parallel to the magnetic edge in the
opposite direction of the classical snake-orbit
[Fig.~\ref{fig:trajectories}(b)].

Now the question arises how this picture is modified if one
includes a hard-wall confinement in the $y$-direction
$V(y>W/2)=\infty$. As argued below, the situation will change
dramatically since the system is no longer translationally
invariant along the $y$ direction and the wave function vanishes
at the wire boundaries. The effect of the weak confinement can be
taken into account qualitatively following the ideas used in band
theory and assuming that the edge mode behaves like a
quasi-particle moving along the magnetic well with effective mass
$m^{*}=1.43m_e$. Namely, expanding the electron energy near the
minimum point $E_0$ for $k_0l_B=-0.767$ in the manner
\begin{equation}\label{quasipart}
  E(k)\rightarrow E_0+ (k-k_0)^2/2m^{*}, \label{eq:approx}
\end{equation}
and presenting the electron wave function as
\begin{equation}
 \Psi(x,y) = f(y)\phi_0(x)
\end{equation}
[where $\phi_0(x)$ is the electron wave function at the minimum
point $k_0$] one can get the effective Schr\"{o}dinger equation
for the envelope function $f(y)$. The Hamiltonian for this
equation is obtained just replacing $k-k_0\rightarrow
-i\partial/\partial y$ in energy expression (\ref{quasipart}) and
adding the confinement potential:
\begin{equation}
  H_0 = E_0 - (1/2m^*)\partial^2/\partial y^2+V(y).
\label{eq:zerohamil}
\end{equation}
In the case of hard wall confinement it immediately leads to the
quantized energy spectrum
\begin{equation}
  E_n = E_0+(n\pi/l_B )^2/2m^*,\label{eq:}
\end{equation}
which is both qualitatively and quantitatively in good agreement
with the exact result shown in Fig. \ref{fig:energy1}, at least
for lower energy values ($|k-k_0|$ small). It is obvious that for
higher $|k-k_0|$ the parabolic approximation (\ref{eq:approx}) is
not valid anymore.

The system studied here can be realized experimentally by growing
a tilted 2DEG with the left and right parts having different
slopes.\cite{leadbeater95} This results in an out-of-plane kink in
the 1D wire. If a (large) homogeneous magnetic field is then
applied, electrons will be attracted towards the kink where the
magnetic field changes abruptly and form a quasi 1D system, which
is confined both electrically and magnetically. Another way to
create the steplike magnetic field in the wire is by growing a
perpendicularly magnetized magnetic stripe on top of the
heterostructure, containing the wire.\cite{nogaret00} The
magnetic field then has a different sign underneath or away from
the stripe.

In summary, we proposed a technique to confine snake-orbit states
and localize them spatially near a magnetic-field discontinuity
(or edge) by introducing an interference between them. These
caged states are bound to the magnetic edge and have an energy
lower than that of the first Landau level. Their number increases
as the magnetic field becomes stronger or as the width of the
wire becomes larger and the confinement decreases. These states
can be understood as an interference between two
counter-propagating snake-orbit states. We also provided a simple
approximation which rendered clear their physical origin. They
have a pure quantum mechanical origin and cannot be understood or
treated classically.

This work was partially supported by the Inter-university
Micro-Electronics Center (IMEC, Leuven), the Flemish Science
Foundation (FWO-Vl), the Belgian Interuniversity Attraction Poles
(IUAP), and the Flemish-Chinese bilateral programme. J.R. was
supported by ``het Vlaams Instituut voor de bevordering van het
Wetenschappelijk \& Technologisch Onderzoek in de Industrie"
(IWT).

\begin{figure}[t]
\caption{The lowest energy values for a quantum wire of width $W$
subjected to a homogeneous magnetic field $B/B_0=500$ (solid
dots), and to a $B$/$-B$ profile (open dots). The inset shows the
wave functions for the ground state ($N=1$) and for the fifth
lowest state $(N=5)$ in case of the $B$/$-B$ profile.}
\label{fig:output}
\end{figure}

\begin{figure}[t]
\caption{The energies for the lowest bound states as function of
the magnetic field strength $B$. The dotted lines correspond to
our approximation [Eq.~(\ref{eq:})], which is only shown for
$E/\hbar \omega_c <0.35$. The thin dashed line corresponds to the
lowest energy snake-orbit state in the absence of confinement.}
 \label{fig:energy1}
\end{figure}

\begin{figure}[t]
\caption{Dispersion relation of electrons subjected to a $B$/$-B$
profile. The thin
solid line separates the different regions A 
and B corresponding, respectively, to states which cannot and can
be understood classically. The bound states with energy below the
first Landau state are represented by the bold line.
The dotted line shows the displaced parabolic approximation with
$k_0/l_B=-0.767$, $E_0/\hbar\omega_c=0.295$, $m^{*}/m_e=1.43$.}
\label{fig:spectrum}
\end{figure}

\begin{figure}[t]
\caption{Schematic trajectories and corresponding wave functions
for the states indicated by the solid dots in
Fig.~\ref{fig:spectrum}, i.e. for $kl_B=-0.3$ and for
$kl_B=-1.35$.}
\label{fig:trajectories}
\end{figure}

\end{document}